\documentclass{aa}
\usepackage{graphicx,times}
\newcommand{\vy}{\object{VY\,CMa} }
\newcommand{\vyp}{\object{VY\,CMa}}
\def\mso{\, M_\odot}
\def\lso{\, L_\odot}

\def\kms{\, {\rm km}\, {\rm s}^{-1}}
\def\simgr{\mathrel{\hbox{\rlap{\hbox{\lower4pt\hbox{$\sim$}}}\hbox{$>$}}}}
\def\msoy{\, \mso~{\rm yr}^{-1}}

\begin{document}
\topmargin0.0cm
\title{Diffraction-limited Speckle-Masking Interferometry of 
the Red Supergiant \vy\thanks{Based on data collected at the European
Southern Observatory, La Silla, Chile}}
\author{M.\ Wittkowski\inst{1} \and N.\ Langer\inst{2} \and 
G.\ Weigelt\inst{1}} 
\institute{Max-Planck-Institut f\"ur Radioastronomie, Auf dem H\"ugel 69,
D-53121 Bonn, Germany
\and
Institut f\"ur Physik, Universit\"at Potsdam, D-14415 Potsdam, Germany}
\thesaurus{1(03.20.2; 08.19.3; 08.12.1; 08.13.2; 08.09.2 VY\,CMa)}
\mail{Markus Wittkowski, E-Mail: mw@speckle.mpifr-bonn.mpg.de}
\date{Received \dots ; accepted \dots}
\maketitle
\begin{abstract}
We present the first diffraction-limited 
images of the  mass-loss envelope of the red supergiant star VY~CMa.
The two-dimensional optical and NIR images were reconstructed from 3.6\,m 
telescope speckle data using bispectrum speckle interferometry. 
At the wavelengths $\sim$\,0.8\,$\mu$m (RG\,780 filter), 1.28\,$\mu$m, and
2.17\,$\mu$m the diffraction-limited resolutions of 46\,mas, 73\,mas, and
124\,mas were achieved. 
All images clearly show that the circumstellar envelope of VY CMa is 
non-spherical.
The RG\,780, 1.28\,$\mu$m, and 2.17\,$\mu$m FWHM Gau\ss~fit diameters are 
67\,mas$\times$83\,mas, 80\,mas$\times$116\,mas and 138\,mas$\times$205\,mas, 
respectively, or $100\,{\rm AU}\times125\,{\rm AU}$,
$120\,{\rm AU}\times174\,{\rm AU}$ and $207\,{\rm AU}\times308\,{\rm AU}$
(for a distance of 1500\,pc).
We discuss several interpretations for the asymmetric morphology.
Combining recent results about the angular 
momentum evolution of red supergiants and their pulsational properties,
we suggest that \vy is an immediate progenitor of \object{IRC\,+10\,420},
a post red supergiant during its transformation into a Wolf-Rayet star.
\end{abstract}
\keywords{Techniques: interferometric -- Stars: supergiants -- Stars: late-type
-- Stars: mass-loss -- 
Stars: individual: VY\,CMa}
\section{Introduction}
\vy (\object{HD\,58061}, \object{SAO\,173591}) is one of the most luminous red 
supergiants in the Galaxy with $L \simeq 4\, 10^5\, \lso$ (Sopka et al. 1995,
Jura \& Kleinmann 1990) and is, therefore, an ideal candidate for the 
study of the progenitor phases of a supernova. 
Its distance is about 1500 pc, its 
spectral type is M5~Iae, and it is variable with a period of $\sim$2200 days
(Jura \& Kleinmann \cite{jura}; Malyuto et al. \cite{malyuto}; Danchi et al. 
\cite{danchi}; Knapp \& Morris \cite{knapp} ; Imai et al. \cite{imai}).
Herbig (\cite{herbig}) found that \vy is embedded 
in an optical nebula with a size of
$8^{\prime\prime}\times 12^{\prime\prime}$ at 650\, nm.
The first high-resolution  observations of the
dust shell of \vy have been reported by
McCarthy \& Low (\cite{mccarthy1}),
McCarthy et al. (\cite{mccarthy2}) and 
Danchi et al. (\cite{danchi}).
\vy is a source of H$_2$O, OH and SiO maser emission 
(e.g. Masheder et al. \cite{masheder},
Bowers et al. \cite{bowers1}, 
Bowers et al. \cite{bowers2},
Imai et al. \cite{imai},
Humphreys et al. \cite{humphreys},
Richards et al. \cite{richards}).
The H$_2$O masers are distributed in an east-west 
direction, whereas OH masers are distributed in a north-south direction, 
possibly
indicating a disk and a polar outflow.
HST FOC images of the envelope of \vy were obtained by Kastner and Weintraub
(\cite{kastner2}). They show an asymmetric flux distribution in an
approximately east-west direction with a brighter core of pure 
scattered light elongated from SE to NW.

In this {\it Letter} we present diffraction-limited $\sim$\,0.8\,$\mu$m,
1.28\,$\mu$m, and 2.17\,$\mu$m bispectrum speckle interferometry observations
of the mass-loss
envelope of \vyp. The observations are presented in Section~2, and
the non-spherical shape of the envelope
is discussed in Section~3. Clues for \vyp 's evolutionary
state are derived in Section~4.  
\section{Observations}
\begin{figure*}
\begin{center}
\resizebox{0.75\hsize}{!}{\includegraphics{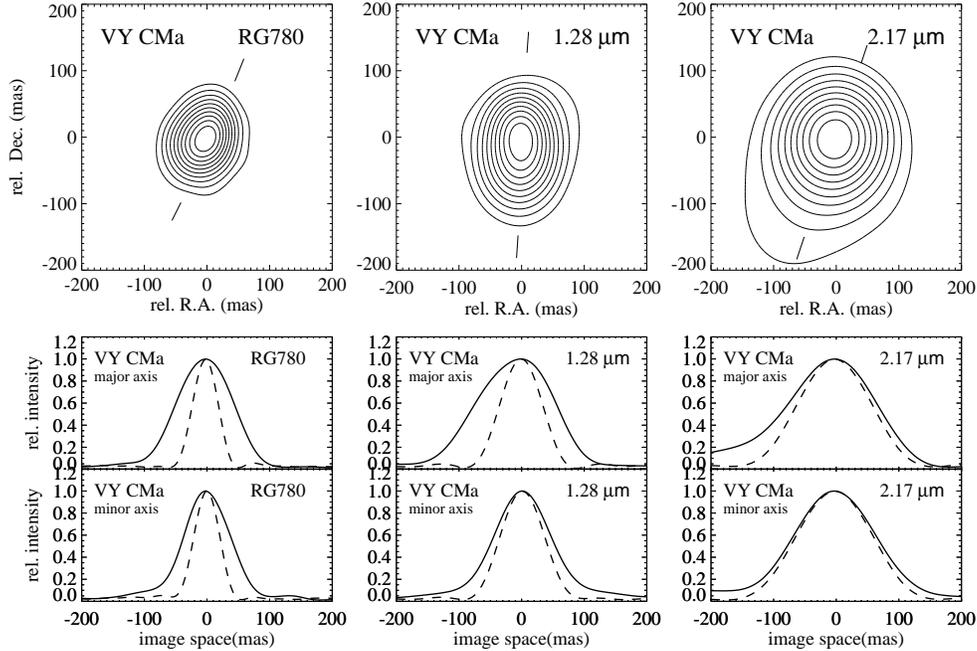}}
\end{center}
\vspace{0.5cm}%
\caption{Top: Contour plots of the RG\,780 (left), 1.28\,$\mu$m (middle) 
and 2.17\,$\mu$m filter (right) bispectrum speckle interferometry 
reconstructions of \vy. The contours are plotted from 15\% to
100\% in 10 steps. North is at the top and east to the left.
Bottom: Intensity cuts through the centers of the reconstructed 
images along the major and minor axes (solid lines). The position angles
of these axes (RG\,780: 153$^\circ$, 1.28\,$\mu$m: 176$^\circ$, 2.17\,$\mu$m:
160$^\circ$) are taken from the two-dimensional fits to the visibility
functions (see Tab. \protect\ref{fits}) and indicated in the contour plots.
The dashed curves are cuts through the reconstructed images of the reference
stars.}
\label{bilder}
\end{figure*}

\begin{figure*}[btp]
\begin{center}
\resizebox{0.3\hsize}{!}{\includegraphics{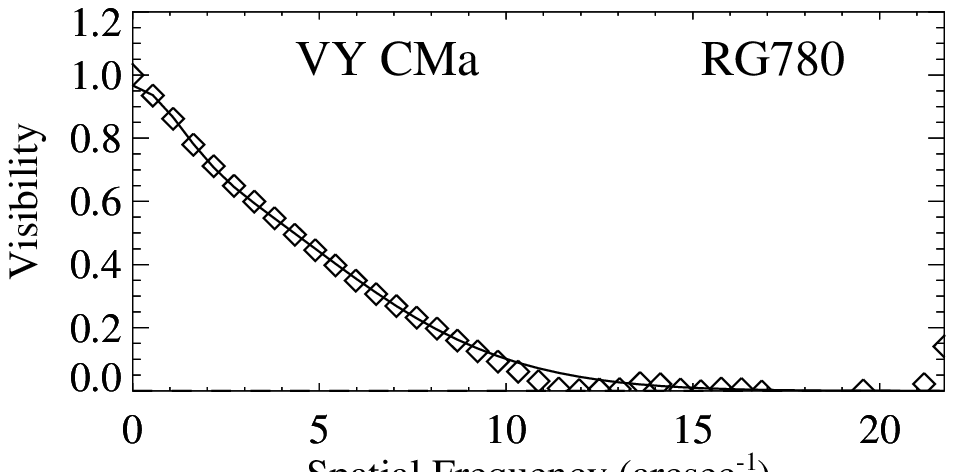}}
\resizebox{0.3\hsize}{!}{\includegraphics{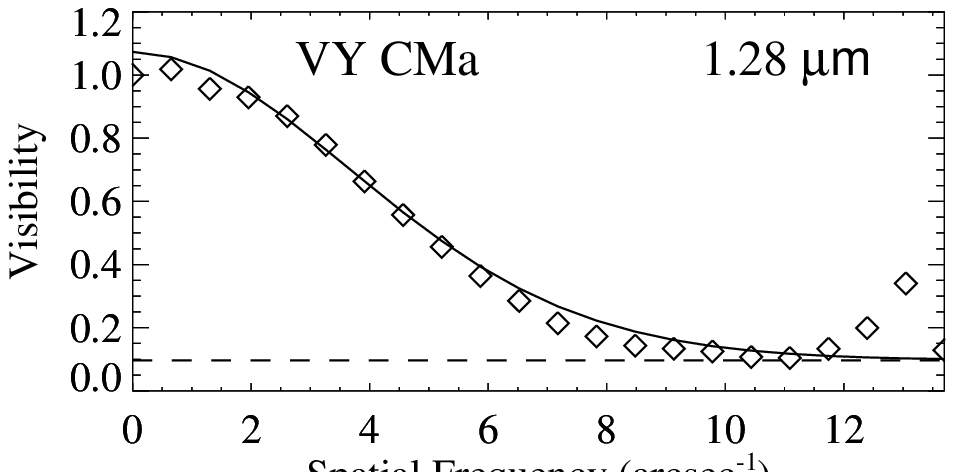}}
\resizebox{0.3\hsize}{!}{\includegraphics{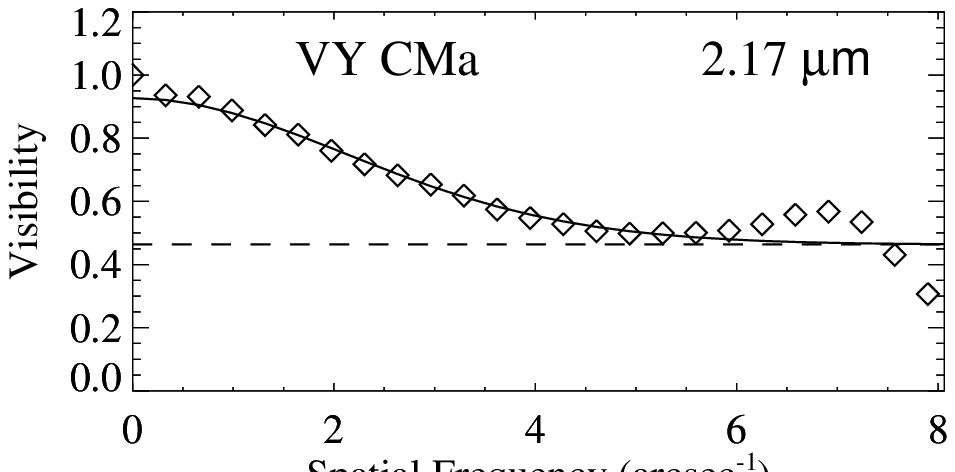}}
\end{center}
\vspace{0.3cm}%
\caption{Azimuthally averaged visibility functions of \vy and of the 
corresponding model flux distribution:
(left) filter RG\,780, (middle) 1.28\,$\mu$m filter and (right) 
2.17\,$\mu$m filter.
The diamonds indicate the observations, the solid line the
azimuthally averaged fit curve of a two-dimensional Gaussian model flux 
distribution plus an unresolved object, and the dashed line the constant 
caused by just the unresolved component.
The RG\,780 visibility function is best described by two Gaussian flux 
distributions.
For the fit parameters see Table \protect\ref{fits}. The visibility
errors are $\pm$ 0.1 up to 80\% of the telescope cut-off frequency and
$\pm$ 0.2 for larger frequencies.}
\label{visis}
\end{figure*}

The optical and NIR speckle interferograms were obtained with 
the ESO 3.6\,m telescope at La Silla on February 6,7 and 8, 1996. The optical 
speckle interferograms were recorded through the edge filter RG\,780 
(center wavelength of the filter/image intensifier combination: 
$\sim$\,0.8\,$\mu$m; effective filter width: $\sim$\,0.07\,$\mu$m) with our
optical speckle camera described by Hofmann et al. (\cite{hofmann}). The near 
infrared speckle interferograms were recorded with our 
NICMOS\,3 camera 
through interference filters with center wavelength/FWHM bandwidth of 
1.28\,$\mu$m/0.012\,$\mu$m and 2.17\,$\mu$m/0.021\,$\mu$m. The observational
parameters (number of frames, exposure time per frame, pixel scale and seeing) 
are listed in Table \ref{obs}. Diffraction-limited images were reconstructed 
from the speckle interferograms by the bispectrum speckle interferometry method 
(Weigelt \cite{weigelt1}; Lohmann et al. \cite{lohmann}; 
Weigelt \cite{weigelt2}). The power spectrum of \vy was determined with the
speckle interferometry method (Labeyrie 1970).
The atmospheric speckle transfer functions were derived from speckle 
interferograms of the unresolved stars H42071 (RG\,780, 2000 frames), 
IRC 30098 (1.28\,$\mu$m, 600 frames) and 1 Pup (2.17\,$\mu$m, 800 frames).\\
\begin{table}[t]
\caption{Observational parameters.}
\begin{tabular}{l l l l l l}
filter & number & exposure & pixel scale & seeing\\ 
       & of frames & time & & \\
\hline
RG\,780     & 2000  & $ 60\, {\rm ms}$ & $7.2\,{\rm mas}  $ & 1\farcs 5\\
1.28\,$\mu$m & 800   & $150\, {\rm ms}$ & $ 23.8\,{\rm mas}$ & 1\farcs 0\\
2.17\,$\mu$m & 800   & $100\, {\rm ms}$ & $ 47.6\,{\rm mas}$ & 1\farcs 5\\ \hline
\end{tabular}
\label{obs}
\end{table}
%
%
Figure \ref{bilder} shows contour plots and intensity cuts of the 
reconstructed bispectrum speckle interferometry images of \vyp. The 
resolutions of the 0.8\,$\mu$m, 1.28\,$\mu$m and 2.17\,$\mu$m images are 
46\,mas, 70\,mas and 111\,mas, respectively. The envelope of \vy is 
asymmetric at each of the three wavelengths.  
The object parameters were determined by 
two-dimensional model fits to the visibility functions. 
The models consist of two-dimensional elliptical Gaussian flux distributions 
plus an additional unresolved component.
The $\sim$\,0.8\,$\mu$m image is best described by two Gaussians while the 
1.28\,$\mu$m and the 2.17\,$\mu$m images are best described by one 
Gaussian and an additional unresolved component. 
The best-fit parameters are listed in Table \ref{fits}.
We present the 
azimuthally averaged visibility function of \vy together with the 
corresponding azimuthally averaged two-dimensional fits in Fig. \ref{visis},
in order to show the wavelength-dependent relative flux contribution
of the unresolved component (dashed line).
\begin{table}[t]
\caption{\vyp 's parameters derived from the model fits to the visibility
functions (one {\em two-dimensional} Gaussian flux distribution plus an 
unresolved object for the 1.28\,$\mu$m and 2.17\,$\mu$m observations and two 
{\em two-dimensional} Gaussian flux distributions for the RG\,780 data). 
The parameters are the position angle of the major axis, 
the axes ratio (major/minor axis), the FWHM of major and minor 
axes, the azimuthally averaged FWHM diameter and the relative flux 
contributions of the Gaussian and the unresolved object. We estimate
the errors of the position angles to $\sim\pm10^\circ$ and those of
the FWHM sizes to $\sim\pm15\%$.
}
\begin{tabular}{l l l l l l}
data set                  &RG\,780        & 1.28\,$\mu$m & 2.17\,$\mu$m \\ 
\hline
PA ($^\circ$) of the major axis & 153/120 & 176       & 160        \\
Axes ratio                & 1.2/1.1       & 1.5       & 1.5        \\
Major axis (mas)          & 83/360        & 116       & 205        \\
Minor axis (mas)          & 67/280        & 80        & 138        \\
Average diameter (mas)    & 74/320        & 96        & 166        \\
rel. flux of Gaussian     & 0.75/0.25     & 0.91      & 0.50       \\
rel. flux of unres. comp. & 0.00          & 0.09      & 0.50       \\
\hline
\end{tabular}
\label{fits}
\end{table}
\section{Interpretation}
Figures 1 and 2 and Tab.~\ref{fits} indicate that \vy 
consists of both an unresolved component and a resolved asymmetric extended 
component with an increasing average diameter at longer wavelengths. 
The unresolved component is probably the star itself or an 
additional compact circumstellar object. The relative intensity of this 
unresolved component decreases towards shorter wavelengths.
The complete obscuration of the unresolved component close to the maximum of 
the stellar spectral energy distribution indicates a high optical depth 
of the circumstellar envelope, in agreement with the value from 
Le~Sidaner and 
Le~Bertre of $\tau_{\rm 10\,\mu m}=2.4$.
The visibility functions, the images and the best fit parameters 
given in Tab.~\ref{fits} at the three different wavelengths
can be used as additional constraints in future two-dimensional radiation 
transfer modeling of the dust distribution.
The resolved structures seen in Fig.~1 belong to the 
circumstellar envelope of \vyp. In fact, the size of the image is of the
order of the dust condensation radius $R_{\rm c}$ which has been
estimated by Le~Sidaner \& Le~Bertre within their spherically symmetric
model to $R_{\rm c} \simeq 12 R_{\star}$ 
with $R_{\star}\sim 4000\,R_\odot$. 
They note, however, that their
model fit to the spectral energy distribution of \vy is not entirely
satisfactory and argue that the envelope of \vy may be non-spherical,
as indicated by the high level of polarisation (Kruszewski \& Coyne 1976).
Danchi et al. (1994) found $R_{\rm c} \simeq 5 R_{\star}$.

Several interpretations for the non-sphericity seen in all three images 
(Fig.\,1) are possible.
The position angle of the major axes of the approximately
elliptical shapes is similar, although
not identical for all three cases (153$^{\circ}$ to 176$^{\circ}$;
see Table~2). 

This position angle is approximately perpendicular to the major axis of the 
H$_2$O maser distribution (Richards et al. \cite{richards}) and 
similar to the distribution of the OH masers (e.g. Masheder et al. 
\cite{masheder}). 
Accordingly, we can interprete the structure of the 
circumstellar envelope of \vy as a bipolar outflow in a north-south direction
caused by an obscuring equatorial disk in an east-west direction. 
The existence of an obscuring disk is supported by the obscuration of the
central star at optical wavelengths.
Such geometry was also discussed for \object{IRC\,+10\,216} by 
Weigelt et al. (1998) and Haniff \& Buscher (1998) and has already been 
proposed for \vy due to maser observations by 
Richards et al. (\cite{richards}).

Furthermore, we can not rule out that the unresolved component consists
of an optically thick torus hiding a close binary, as was proposed for the 
Red\,Rectangle (e.g. Men'shchikov et al. 1998). This could lead to
an asymmetric outflow in a north-south direction.

The mass-loss mechanism of \vy could also be erratic or stochastic,
similar to the clumpy pulsation and dust-driven mass-loss events
recently detected in the prototype carbon star \object{IRC~+10\,216} 
(see Weigelt et al. 1998, Haniff \& Buscher 1998).
Although the reason for this anisotropy is unknown, and the physics of 
mass-loss in oxygen-rich red supergiants may differ from those in 
carbon-rich stars,  the common properties of \vy and \object{IRC~+10\,216}
--- both are pulsating cool luminous stars with extended convective 
envelopes --- could result in similar mass-loss features. 
But individual clumps are not observable because of the larger distance of
\vyp, which leads to a regular asymmetric image elongated 
in a north-south direction. 

Finally, our results are also conceivable with a more regular geometry 
of \vyp 's circumstellar envelope, for example, a disk-like envelope which 
appears elongated in a north-south direction due to the projection angle 
(discussed in Sec.\,4 in more detail), which was proposed earlier for \vy 
on the basis of optical, infrared (Herbig 1970, 1972; McCarthy 1979; 
Efstathiou \& Rowan-Robinson 1990), and maser observations 
(van Blerkom 1978, Morris \& Bowers 1980, Zhen-pu \& Kaifu 1984).
\section{Evolutionary status and conclusions}
With a distance of $\sim$\,1500\,pc, the luminosity of \vy amounts
to $\sim 4\, 10^5\lso$ (see Jura \& Kleinmann 1990). 
Figure~3 compares this luminosity to a stellar evolutionary track
of a 40$\mso$ star in the HR diagram (see Langer 1991, for details
of the computational method). It constrains the initial mass of \vy
to the range $30 ... 40\mso$, in agreement with e.g. Meynet et al. 
(1994),  although models with rotation --- which are not yet available
for the post main sequence evolution at this mass ---
may lead to somewhat smaller values (cf. Heger et al. 1997). 
Accordingly, it is likely that \vy will
transform into a Wolf-Rayet star during its further evolution.
This is supported by the very high 
observed mass-loss rate of \vy of $\sim 10^{-4}\msoy$ (Jura \& Kleinmann 1990).
\begin{figure}[t]
\resizebox{0.85\hsize}{!}{\includegraphics{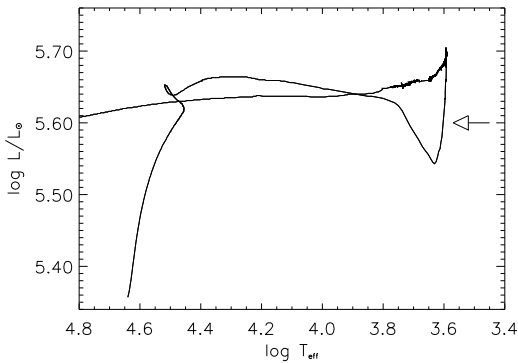}}
\vspace{0.3cm}%
\caption{
Evolutionary track of a 40$\mso$ star of solar metallicity, from the
zero age main sequence to the red supergiant $\rightarrow$ Wolf-Rayet star
transition. The luminosity of VY CMa is indicated by an arrow.
}
\end{figure}
In order to obtain a disk-like structure (see Sec.\,3),
the most likely mechanisms
involve angular momentum. 
There is no indication of a binary companion
to VY~CMa, either from the spectrum of VY~CMa (Humphreys et al. 1972) or
from high resolution imaging (see above). While this only excludes
massive companions, it appears to be viable 
that the axial geometry is due to the star's
rotation.
Direct evidence for rotation being capable of producing a disk-like
structure around red supergiants --- possibly through the
Bjorkman-Cassinelli-mechanism of wind compression (see Ignace et al. 1996) ---
comes from bipolar structures found in the AGB star
\object{V~Hydrae} (Stanek et al. 1995), for which rapid rotation
($v \sin i \simeq 13\kms$) has been
directly inferred from photospheric line broadening (Barnbaum et al. 1995).

According to Heger \& Langer (1998), red supergiants drastically increase 
their surface rotation rate shortly before and during
the evolution off the Hayashi line. Therefore, 
a disk-like envelope surrounding \vy
may indicate that such a spin-up is currently in progress.
As strong mass-loss from a convective star 
acts as a spin-down mechanism (Langer 1998, Heger \& Langer 1998), 
a red supergiant must previously have lost the major part of its envelope 
for the competing spin-up process to dominate. 
This strongly supports the argument 
that the remaining mass of \vyp 's convective envelope
is in fact small, and that \vy is just about to leave the Hayashi line. 
It is also consistent with the very long observed 
pulsation period of \vy of about $6\,$yr
according to the pulsational analysis of Heger et al. (1997),
who showed that such large periods can be obtained in red supergiants 
for small envelope masses (cf. their Fig.~2a).

With this interpretation, \vy 
represents the immediate progenitor state of \object{IRC\,+10\,420}, 
currently a mid A~type supergiant evolving bluewards on human time scales 
on its way from the red supergiant stage 
to become a Wolf-Rayet star (Jones et al. 1993, Kastner \& Weintraub 1995). 
A comparison with the 40$\mso$ model shown in Fig.~3
predicts a current mass of \vy of ~15$\mso$ and a surface helium mass fraction
of $Y\simeq 0.40$.  

It is remarkable in the
present context that bipolar outflows are observed in \object{IRC\,+10\,420}
(Oudmaijer et al. 1994, 1996, Humphreys et al. 1997).
A disk-like structure of VY~CMa's envelope could be the basis for
such flows, which occur when a fast wind originating from the star 
in a post-red supergiant stage interacts with a previously formed disk,
according to hydrodynamic simulations of interacting wind flows
(e.g., Mellema 1997, Garc\'{\i}a-Segura et al. 1998).
\begin{acknowledgements}
  This work has been
  supported by the Deutsche Forschungsgemeinschaft through grants
  La~587/15-1 and 16-1.
\end{acknowledgements}
\end{document}